%% file: main.tex
\documentclass[conference,a4paper]{IEEEtran}
\IEEEoverridecommandlockouts

\usepackage{cite}
\usepackage{url}
\usepackage[inline]{enumitem}
\usepackage{grffile}
\usepackage[ruled,vlined]{algorithm2e}

\usepackage{amsmath,amssymb,amsfonts}

\usepackage[utf8]{inputenc}

\usepackage[english]{babel}

\usepackage{tabularx}
\usepackage[utf8]{inputenc}
\usepackage{lipsum}
\usepackage{float}
\usepackage{blindtext}

\usepackage{nomencl}

\SetCommentSty{mycommfont}
\usepackage{textcomp}
\usepackage{xcolor}
\usepackage{lipsum}
\usepackage{xfrac}
\usepackage{commath}
\usepackage{flexisym}
\usepackage{makecell}
\usepackage{mathtools}
\usepackage{multicol}
\usepackage[labelfont=bf]{caption} 
\usepackage{caption}
\usepackage[section]{placeins}
\usepackage{cleveref}
\usepackage{float}

\def\BibTeX{{\rm B\kern-.05em{\sc i\kern-.025em b}\kern-.08em
    T\kern-.1667em\lower.7ex\hbox{E}\kern-.125emX}}

\SetKwComment{Comment}{$\triangleright$\ }{}

\usepackage{amsthm}

\newtheorem{lemma}{Lemma}

\newtheorem{theorem}{Theorem}

\makeatletter
\newcommand{\removelatexerror}{\let\@latex@error\@gobble}
\makeatother

\begin{document}

\title{BunchBFT: Across-Cluster Consensus Protocol}
\author{\IEEEauthorblockN{Salem Alqahtani and Murat Demirbas}
salemmoh,demirbas@buffalo.edu}
\maketitle

\input{Abstract}
\input{Introduction}
\input{Related}
\input{Preliminaries}
\input{Protocol}
\input{Correctness}
\input{Evaluation}
\input{Conclusions}

\bibliographystyle{IEEEtran}
\bibliography{Survey.bib}
\end{document}

%% file: Abstract.tex
\begin{abstract}

In this paper, we present BunchBFT Byzantine fault-tolerant state-machine replication for high performance and scalability. At the heart of BunchBFT is a novel design called the cluster-based approach that divides the replicas into clusters of replicas. By combining this cluster-based approach with hierarchical communications across clusters, piggybacking techniques for sending messages across clusters, and decentralized leader election for each cluster, BunchBFT achieves high performance and scalability.

We also prove that BunchBFT satisfies the basic safety and liveness properties of Byzantine consensus. We implemented a prototype of BunchBFT in our PaxiBFT framework to show that the BunchBFT can improve the MirBFT's throughput by 10x, depending on the available bandwidth on wide-area links.

\end{abstract}

%% file: Introduction.tex
\section{Introduction}

The emergence of blockchains kick started a renewed interest in Byzantine fault-tolerant(BFT) protocols. Tendermint~\cite{Tendermint} implemented a PBFT~\cite{PBFT} like protocol for permisssioned blockchain systems. Later, many BFT consensus protocols emerged to improve communication efficiency including Hotstuff~\cite{Hotstuff}, Fast-Hotstuff~\cite{FHS}, Bidl~\cite{qi2021bidl}, Kauri~\cite{neiheiser2021kauri}, and SBFT~\cite{golan2018sbft}. Unfortunately, the protocols still suffer from scalability limitations because of a single leader coordination~\cite{alqahtani2021bottlenecks,yang2018linbft}.

A recent effort to overcome the single leader bottleneck by allowing multiple parallel leaders~\cite{Mir,gupta2021rcc,BigBFT} to reduce the disproportionate workload on the leader and the number of messages sent/received for one consensus instance. Unfortunately, those protocols inherit PBFT's message complexity, lack pipelining support, or depend on a view-change protocol to replace the leaders. The performance of the systems degrades severely, especially in geo-distributed settings.

In BFT-based systems, the efficiency of the whole system typically degrades with more nodes joining the network. In this paper, we present the BunchBFT protocol to address these shortcomings. In BunchBFT, we are able to boost scalability and throughput and address latency problems. By sharding the round, clustering nodes to boost concurrency and fault containment in each cluster, and decentralizing the leader election by offloading the leader election from the system's core to its edge, we spread out the network workload so that node is not responsible for processing the entire network's transactional load. Instead, each node only maintains information related to its cluster. Eventually, the information contained in a cluster can still be shared among other nodes, which keeps the ledger decentralized and secure because every node can still see all the ledger entries.
To address network bandwidth shortages on wide-area links, BunchBFT-X was built to save on both the size and number of the communications. BunchBFT-X introduces a hierarchical communication design to reduce the number of communications and reduces the size of communications by sending a hash of data blocks instead of actual data across clusters.

We provide safety proof of BunchBFT to satisfy correctness. We describe our framework called PaxiBFT which captures our design choices for implementing, benchmarking, and evaluating BunchBFT and BunchBFT-X performance under identical conditions with MirBFT.

Finally, we present experimental evaluations to show that the throughput of BunchBFT is a factor of magnitude better than the MirBFT protocol\cite{Mir}. As the number of nodes increases, BunchBFT scales to process over 300,000 transactions per second. This makes BunchBFT an ideal choice for blockchain systems assuming 128K (block size) and 128 bytes (transaction size).

%% file: Related.tex
\section{Related Work}
\label{sec:BG}

BFT protocols are widely used in permissioned blockchains~\cite{HyperLedger}. Most of these BFT protocols are categorized as single(PBFT~\cite{PBFT}) and multi-leader(MirBFT~\cite{Mir}) systems.

\textbf{PBFT}~\cite{PBFT} provides the first practical solution to the Byzantine problem~\cite{lamport2019byzantine}. PBFT employs an optimal bound of $\!N\!\geq\!\!3F\!+\!1\!$ replicas, where the Byzantine adversaries can only control up to $\!F\!$ replicas. PBFT uses encrypted messages to prevent spoofing and replay attacks, as well as to detect corrupted messages. PBFT employs a leader-based paradigm, guarantees safety in an asynchronous model, and guarantees liveness in a partially synchronous model. PBFT requires $O(n^2)$ in its best case and $O(n^4)$ in the worst case.

\begin{figure}
	\centering
	\vspace*{-6mm}
	\includegraphics[width=3.5in,height=1.1in]{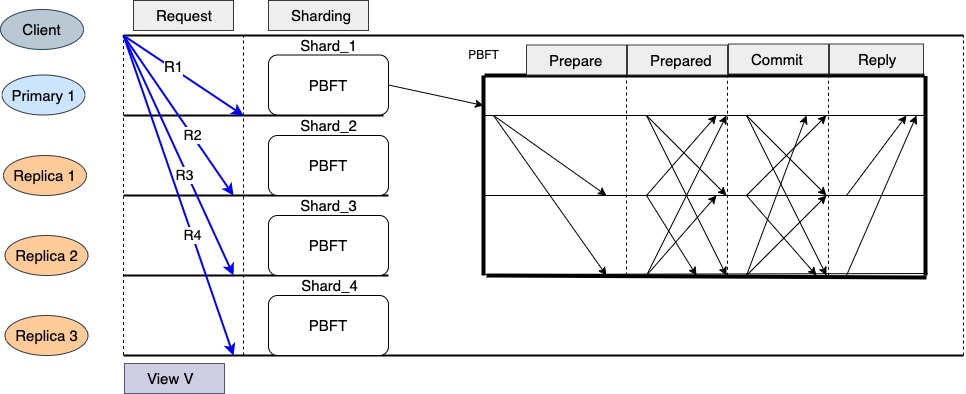}
	\caption{\textbf{MirBFT Protocol}}
	\vspace*{-6mm}
	\label{mir}
\end{figure}

\textbf{MirBFT}~\cite{Mir} is a multi-leader consensus protocol. As illustrated in Figure~\ref{mir}, MirBFT starts by partitioning the request hash space among all leaders to solve duplication attacks and rotates the request hash space among all leaders to solve censorship attacks. It also uses batching and watermarks to facilitate concurrent proposals of batches by multiple parallel leaders. MirBFT proceeds in epochs and each epoch has a single primary and a set of leaders. Each leader will run an independent instance of PBFT~\cite{PBFT}.

%% file: Preliminaries.tex
\section{System Model and Cryptographic Primitives}
\label{sec:back}
\addtolength{\topmargin}{0.01in}
We consider a BFT protocol with N=3F+1 replicas in each cluster. A cluster has a dedicated leader that is responsible for proposing values. The leader in each cluster is elected by accumulating client votes and a replica with the highest number of votes wins the election. We also assume that each node has an ID, which is known to all replicas. The correct replicas follow their specifications while Byzantine replicas are controlled by an adversary and behave arbitrarily.

Following common practice in the literature, we assume a partial synchrony communication model in cluster communication~\cite{dwork1988consensus}, as most BFT protocols of the same kind~\cite{PBFT, Streamlet,golan2018sbft, Hotstuff}, where there is a known network delay bound $\delta$ that will hold after an unknown Global Stabilization Time(GST). After GST, all messages between honest leaders will arrive within time $\delta$. Although we assume partial synchrony, the protocol achieves consistency(i.e., safety) regardless of how long the message delays are or how badly the network might be partitioned. Also, the protocol maintains responsiveness which proceeds as the network delivers~\cite{Hotstuff,hu2021don}.

We take advantage of existing cryptographic tools that are available. Similar to many BFT protocols~\cite{Hotstuff,golan2018sbft,yang2018linbft}, we assume standard digital signatures and public-key infrastructure (PKI) that identify all leader and client processes. The BunchBFT's message exchange patterns are combined with some cryptographic primitives to create digital signatures. BunchBFT uses message digest to detect corrupted messages. Both clients and leaders must be able to verify other leaders' public keys and messages. We also assume a cryptographic hash function $H(.)$ that maps arbitrary input to a fixed size output in a collision-resistant manner.

%% file: Protocol.tex
\section{BunchBFT}
\label{sec:Protocol}

\begin{figure}
  \centering
  \vspace*{-5mm}
  \includegraphics[width=3.5in,height=0.9in]{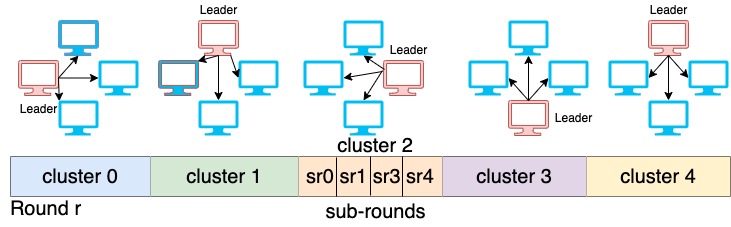}
  \caption{\textbf{Round(r) and sub-round(sr) of BunchBFT}}
  \vspace*{-6mm}
  \label{rounds}
\end{figure}

The BunchBFT executes in rounds(r) as illustrated in~\Cref{rounds}. Each round r is sharded in multiple clusters(cluster0,cluster1..cluster4). Each cluster has a collection of replicas with a dedicated leader where the leader proposes the client's requests to its cluster in sub-rounds(sr0,sr1..sr4) and communicates with other clusters to reach consensus on sub-rounds values. The sub-round length in each cluster should be K where $K>F$ to the fault containment of the leader in each cluster. Specifically, upon F leaders' failure, a new leader can be elected in the cluster without waiting for the increment of the round across clusters.

\begin{figure}
  \centering
  \vspace*{-4mm}
  \includegraphics[width=3.5in,height=1.1in]{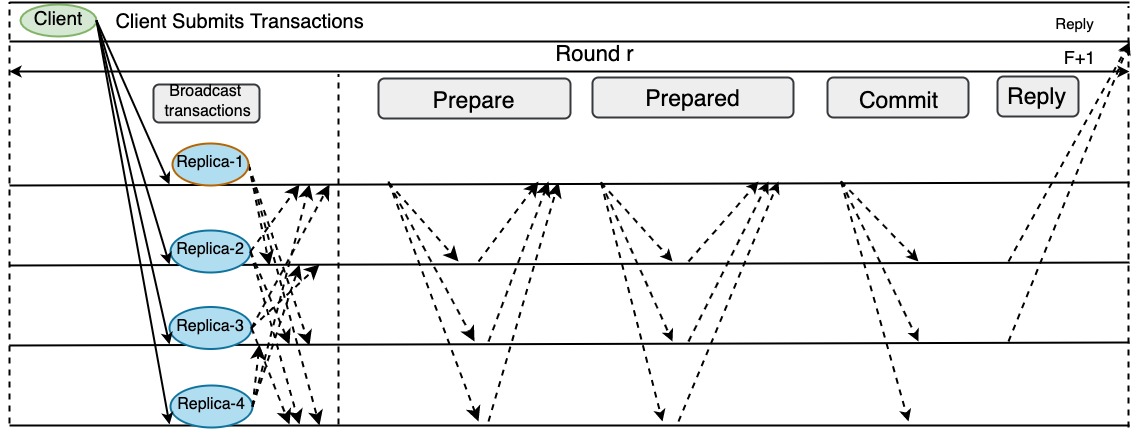}
  \caption{\textbf{Basic BunchBFT Protocol}}
  \vspace*{-6mm}
  \label{basic}
\end{figure}

The BunchBFT divides into basic, pipelined, and BunchBFT-X protocols that we described next.

 
\subsection{BunchBFT Normal Case}
\label{sec:Basic}
\textbf{1- Basic BunchBFT:} The design as described below communicates with nodes from the same cluster to reach consensus assuming no dependencies across clusters.

\begin{algorithm}[hbt!]
  \caption{Broadcast in Replica(i)}
  \label{alg0}
  \DontPrintSemicolon
\For{all replicas} {
 \If{Upon receiving Tx from client}{
     \If{Tx == valid}{
         \If{s == committed}{
            send(Reply)\;
          }
          \Else{
            \If{Lr == replica(i)}{
                reqn++, broadcast(Tx, reqn, H(Tx))\;
            }
            \Else{
              broadcast(Tx, H(Tx))\;
            }
  }
      }      
      \Else{
          \Return reject
          }
 }} 
\end{algorithm}

 \begin{algorithm}[hbt!]
  \caption{Basic BunchBFT in Replica(i)}
  \label{alg1}
\DontPrintSemicolon
 \For{all replicas} {
 \If{Upon receiving F+1 copies of Tx}{
     \text{Block} = \text{Block} $\cup$ Tx
 }
  \Comment*[l]{Prepare Phase}
 \If{replica has max(votes) in Block}{
     \text{replica i } $:=$ \text{Leader(Block)}, \text{hb} $=$ \text{H(Block)}\;
     \If{replica already the leader for r}{
         broadcast(prepare, r, r.sr+1, hb)\;
     }
     \Else{
          broadcast(prepare, r, r.sr=0, hb, H-NB)\;
     }
 }
   \Comment*[l]{Prepared Phase}
   \If{Upon receiving N-F VoteOk.prepare msgs}{
   $\set{\sigma} = Sign_i( \bigcup_{j=1}^{N-F} hb_{j} )$\;
     broadcast(prepared, $\set{\sigma,hb,r,r.sr}$)\;
     }
     \Comment*[l]{Commit Phase}
   \If{Upon receiving N-F VoteOk.prepared msgs}{
     $\set{\sigma} = Sign_i(\bigcup_{j=1}^{N-F} hb_{j})$\;
      broadcast(commit, $\set{\sigma,hb,r,r.sr}$)\;
      send(Reply)\;
       \If{len(sr) == max size of $cluster_{i}$}{
         r=r+1\;
     }
     }
   \If{replica $==$ follower(i)}{
   \If{Upon receiving a prepare msg}{
     send(PrepareVote)\;
     }
     \If{Upon receiving a prepared msg}{
     send(PreparedVote)\;
  }
  \If{Upon receiving a Commit msg}{
     $Block_{i}$ = committed\;
     send(Reply)\;
  }
  }
 }
\end{algorithm}

\begin{figure*}[h]
  \centering
  \vspace*{-4mm}
  \includegraphics[width=\textwidth,height=3.5cm]{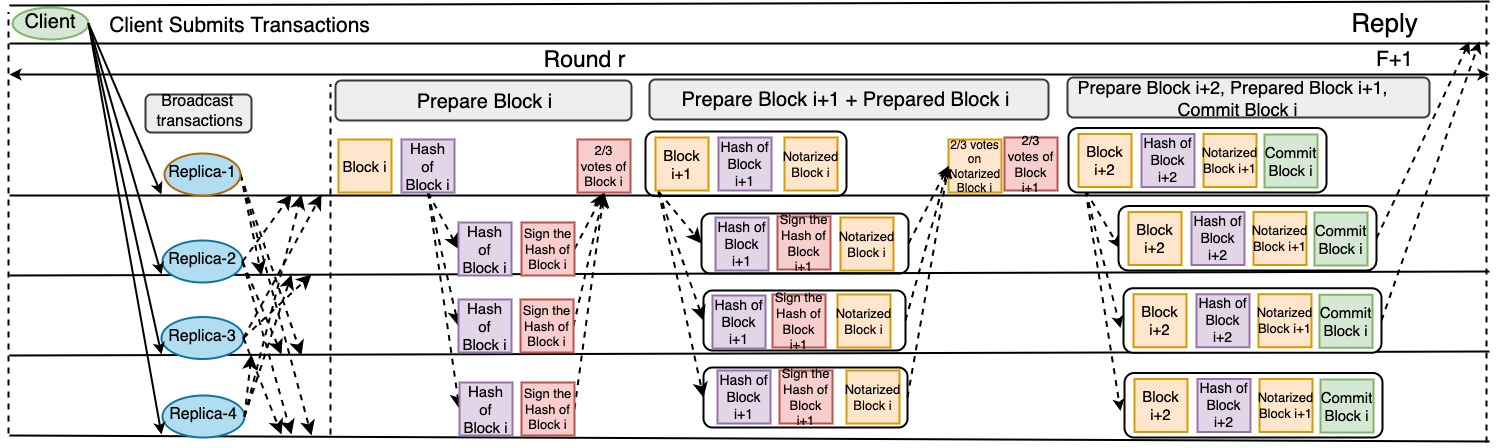}
  \caption{\textbf{Pipelined BunchBFT Protocol}}
  \vspace*{-4mm}
  \label{PipelinedNewBigBFT}
\end{figure*} 

\begin{enumerate}

  \item The client broadcasts a \Big \langle Request op, c, s, Lr \Big \rangle message to the replicas, where op is the write/read operation, c is the client-id, s is the request-number, and Lr is the leader-id that the client prefers for its request. Upon receiving the request, the replica(i) as shown in~\Cref{basic} checks the client table to see if the request-number s has already been executed. If the request was executed previously, replicas reply back to the client with the result. Otherwise, based on the clients' embedded votes, the leader assigns a new request number reqn+1 and re-broadcasts the request. Other replicas also rebroadcast the request(reqn) without increasing as described in~\Cref{alg0}.
  
  \item Based on the clients' votes and the replica's ranking, a single replica(leader) eventually bundles hashes for each request that has at least F+1 copies into a block. The transactions in the block are ordered based on replicas IDs. Upon successfully forming the block and electing the leader, the leader broadcasts the hash of the block to the cluster in a prepare message, \Big\langle Prepare r, sr, H(Block), H-NB\Big\rangle, where r is the latest round number, sr is the sub-round number, H(Block) is the hash of the block, and H-NB is the highest notarized block known to the replica(i). 
  
  \item At the beginning of each round, the leader increases the round number by r+1; otherwise, it increases sr+1. The replicas cannot accept a new prepare message without proof that the leader is extending the last notarized block the replica knows. The proof is to include the H-NB of the previous block in the next block.
  
  \item Upon receiving the prepare message, the H-NB in the prepare message should be the same as the last notarized block in each local replica's log. Upon success, followers vote VoteOk.prepare back to the leader, \Big\langle sr, r, $sign_{i}(H(Block))$\Big\rangle, where $sign_{i}(H(Block))$ is the partial signature of the H(Block). If the leader receives N-F VoteOk.prepare messages, the leader combines sign messages and creates a notarized block $\delta(block)$. The leader broadcasts the notarized block in the prepared messages to all replicas in the same cluster \Big\langle Prepared r, sr, H(Block), NB\Big\rangle where NB is a notarized block. Upon receiving a NB in the cluster, each replica appends it to the log and replies to the leader, \Big \langle sr, r, $sign_{i}$(block(NB))\Big\rangle, where $sign_{i}$(NB) is the partial signature of the NB.
  \item Upon receiving N-F VoteOk.Prepared messages, the leader commits the value and broadcasts commit messages to all replicas where each replica replies to the client. 
\end{enumerate}

\begin{figure*}[h]
  \centering
  \vspace*{-5mm}
  \includegraphics[width=\textwidth,height=3.5cm]{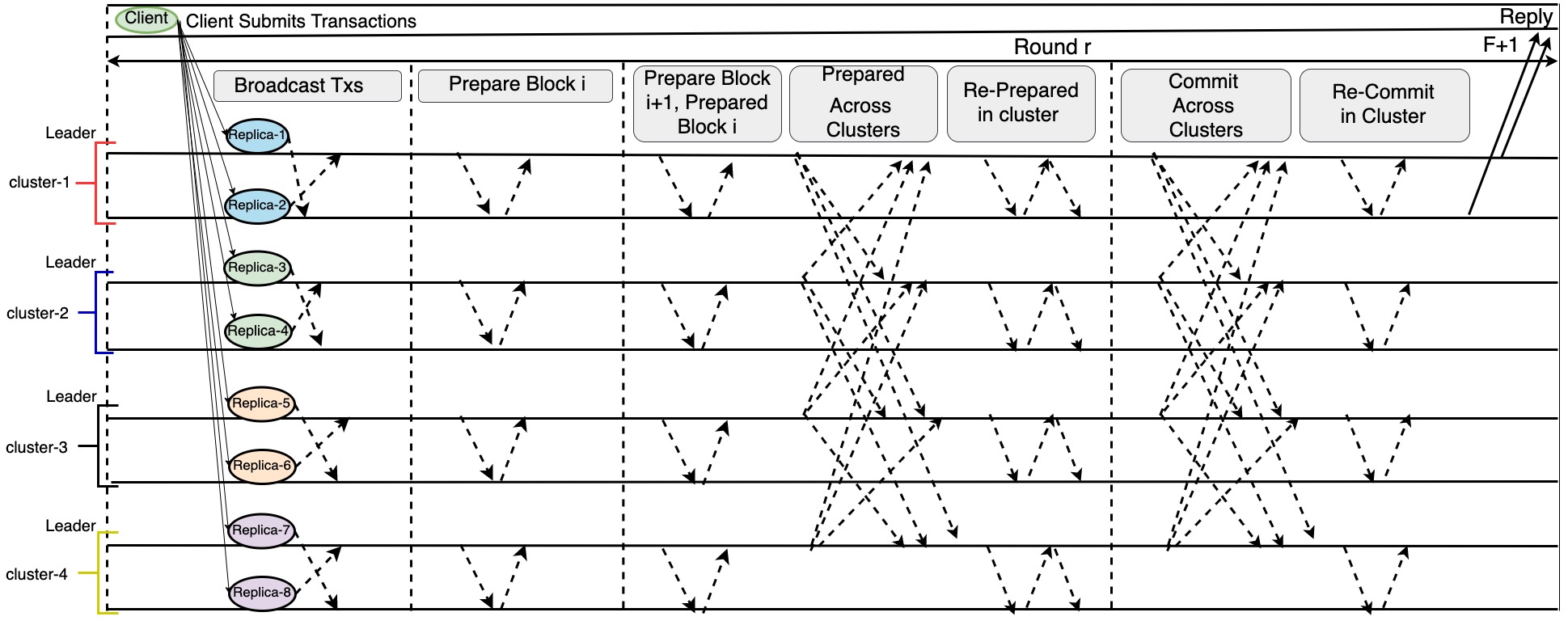}
  \caption{\textbf{BunchBFT-X Protocol}}
  \vspace*{-5mm}
  \label{GeoNewBigBFT}
\end{figure*} 

\textbf{2- Pipelined BunchBFT:} Pipelined BunchBFT extends the basic BunchBFT by applying the pipeline techniques~\cite{Hotstuff,BigBFT} in BunchBFT's phases as illustrated in~\Cref{PipelinedNewBigBFT}. The technique is especially effective when most concurrent transactions do not conflict because their phases consume different resources. Unlike basic BunchBFT where a new consensus instance does not start until the previous consensus instance decides, pipelined BunchBFT starts all phases at the same time from different pipeline sub-round leveraging the network bandwidth and the CPU resources. Upon elected the leader $L_{i}$ for round $r$, the leader pipelines prepare, prepared, and commit phases. The leader $L_{i}$ prepares $block_{i+2}$, prepared $block_{i+1}$, and commit $block_{i}$. \Cref{PipelinedNewBigBFT} illustrates 3 blocks pipeline where each replica of BunchBFT starts a new sub-round. The~\Cref{alg2} shows how BunchBFT pipelines the three phases.

\begin{algorithm}[hbt!]
  \caption{Pipelined BunchBFT in Replica(i)}
  \label{alg2}
\DontPrintSemicolon
  \Comment*[l]{Prepare Phase}   
  \If{replica(i) == $L_{i}$ for r}{
      broadcast([prepare block$_{i}$])\;
       \Comment*[l]{Prepare + Prepared Phase}  
   \If{Upon receiving N-F (VoteOk.prepare(block$_{i}$)) and new Block$_{i+1}$}{
      broadcast([prepare block$_{i+1}$],[prepared block$_{i}$])\;
      }
        \Comment*[l]{Prepare + Prepared + Commit Phase}  
   \If{Upon receiving N-F (VoteOk.prepare(block$_{i+1}$)+VoteOk.prepared(block$_{i}$)) and new Block$_{i+2}$}{
      broadcast([prepare block$_{i+2}$],[prepared block$_{i+1}$],[commit block$_{i}$])\;
      }
     }
  \vspace*{-2mm}
\end{algorithm}
 
\begin{figure}
  \centering
  \vspace*{-3mm}
  \includegraphics[width=3.5in]{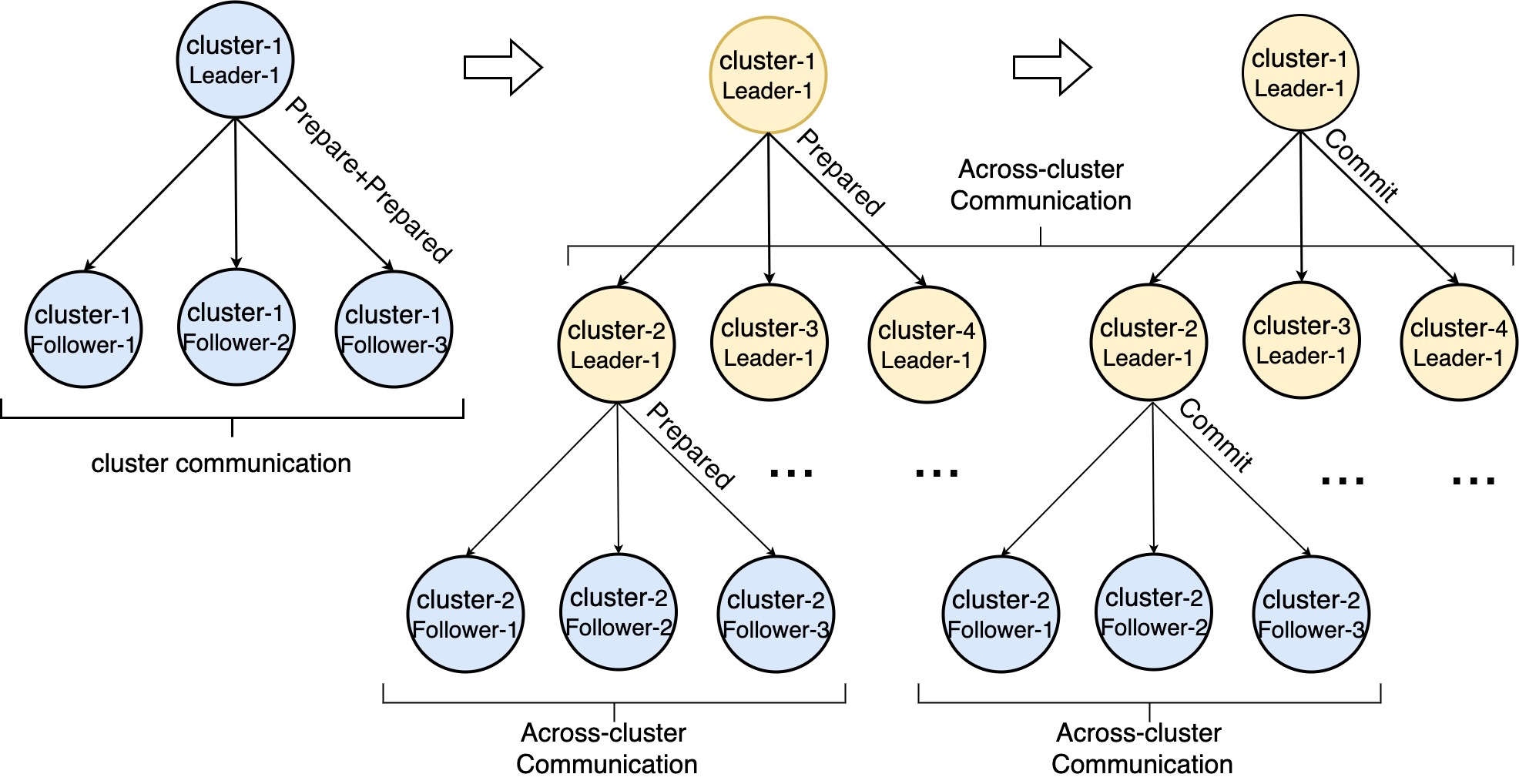}
  \caption{\textbf{BunchBFT-X Communication Pattern}}
  \vspace*{-6mm}
  \label{LeaderXBunchBFT}
\end{figure}

\textbf{3- BunchBFT-X:} BunchBFT-X extends the pipelined BunchBFT by using the piggybacking techniques~\cite{BigBFT} on pipelined BunchBFT. The technique maintains fewer communication messages across clusters when many messages are combined in one message to reduce the number of messages in the network. The design as described below communicates with different clusters to reach a consensus after local communications.~\Cref{GeoNewBigBFT} illustrates 4 clusters where each cluster starts communicating locally and across clusters by sending prepared and commit messages. In~\Cref{LeaderXBunchBFT}, we illustrate the sending process for leaders in and across clusters. The~\Cref{alg3} shows how BunchBFT-X piggybacks the protocol messages.

\begin{algorithm}[hbt!]
  \caption{BunchBFT-X}
  \label{alg3}
\DontPrintSemicolon
\For{ all leaders in each cluster}{
 \If{ Upon agreement on $sr_{i-K}$ in-cluster}{
     \Comment*[l]{send Prepared message across-cluster: NB for $sr_{i-K}$}
     broadcast(NB for $sr_{i-K}$,r)\;
     }
    \Else{
          \Return go to next leader
     }
 \If{Upon receiving Prepared message from each cluster}{
    \Comment*[l]{ Broadcast Prepared message in-cluster}
     broadcast(r, sr, NB for $sr_{i-K}$, H-NB)\;
 }
   \If{N-F VoteOk.Prepared received}{
   $\set{\sigma} = \bigcup_{i=1}^{N-F} Sign_i(NB_{j})$\;
    \Comment*[l]{send Commit message across-cluster: NB for $sr_{i-K}$}
     broadcast($\set{\sigma}$,NB,r)\;
     }
   \If{N-1 VoteOk.Commit received}{
   \Comment*[l]{Broadcast Commit message in-cluster}
     broadcast($\set{\sigma}$,NB,r)\;
     \Comment*[l]{Reply to Client}
     $send(Reply) $\;
     }
  }
   \vspace*{-2mm}
\end{algorithm}

\begin{enumerate}
\item For each cluster, the leader $L_{i}$ in round $r$ piggybacks notarized blocks(NB) for all $sr_{i-K}$ in a single message by sending all $sr_{i-K}$ prepared messages to all other leaders in different clusters with the same ID. Upon receiving the prepared message in other clusters, the leader $L_{j}$ broadcasts the message for all replicas in the same cluster in a local prepared message. Replicas will verify the message and respond back to the $L_{j}$. 
\item Upon receiving N-F votes, $L_{j}$ combines the votes in a commit message. Then, $L_{j}$ sends a commit message both locally for all nodes and across cluster nodes with the same ID. For every leader in different clusters, upon receiving the total number of cluster commit messages, leaders commit locally and reply to the client. 
\end{enumerate}

\subsection{BunchBFT Failure Cases}

Upon failure of a normal case, the BunchBFT revokes the recovery mechanism by switching to a new round/sub-round. If a timeout expires without communication from the leader, the replicas carry out a round change to switch to a new leader. The round/sub-round change in BunchBFT is based on the client's votes and replica ranking between its peers. We described the failure cases below.

\begin{enumerate}

   \item If an elected leader fails in basic and pipelined BunchBFT as was described in section~\ref{sec:Basic}, the N-F honest replicas wait for a timeout to expire. Then, each replica will send a sub-round-change message \Big\langle sr+1, r, H-NB\Big\rangle to the next leader who has the second most votes from the client for the block. Upon receiving the N-F sub-round-change, the newly elected leader prepares the current notarized block or creates a new block with H-NB.

   \item In BunchBFT-X, if an elected leader in cluster-1 did not receive N commit messages, the N-F honest replicas will wait for a timeout to expire. Then, each replica will send a sub-round-change message \Big\langle sr+1, r, H-NB\Big\rangle to the next leader who has the second most votes from the client for the block. The new leader will communicate with the same ID nodes in the different clusters to commit the value. Leaders should communicate across clusters to commit the NB from the previous leader or its block. 

\end{enumerate}

%% file: Correctness.tex
\section{BunchBFT Correctness}
\label{sec:proof}

\textbf{Safety:} BunchBFT guarantees safety in any circumstances even when there is a network delay or network partitions.

\begin{lemma}
For all $F \geq 0$, any two sets of replicas with voting power at least equal to 2F+1 have at least one correct replica in common.
\end{lemma}
\begin{proof}
As the total voting power equals N=3F+1, we have two sets where each has 2F+1 voting power. This means that the intersection of two sets contains at least F+1 voting power in common; therefore, at least one correct replica in both sets.
\end{proof}

\begin{lemma}
In the execution of rounds, the last round is a stable round.
\end{lemma}
\begin{proof}
Assume by contradiction that the last round is not a stable round.\;
1. In a happy path, there are N-F replicas that have the last round stable – a contradiction. 2. Upon failure, the N-F replicas have not successfully completed the new round, hence the new round needs a round-change to be both last and stable round – a contradiction.
\end{proof} 

\begin{lemma}
 If at least one correct leader in a cluster has received global commits, $B_j$ in $sub\!-\!round(sr_{i})$ in $round(r)$, then every other cluster will eventually commit the block $B_j$ in $sub\!-\!round(sr_{i})$ in $round(r)$
\end{lemma}
\begin{proof}
From Lemma 1, we know that a unique block is notarized. Therefore, in each cluster, the block $B_{j}$ in $cluster_{i}$ is committed at $sr_{i}$ in r if each leader receives global commit messages from other clusters to commit the block.
\end{proof} 

\textbf{Liveness:} In BFT protocols, safety and liveness are both guaranteed only in a particularly synchronous environment~\cite{FLP}. Therefore, we show that value will be chosen provided that the network is stable, the delay time is bounded, and there are $N>=2F+1$ correct replicas and less than $F\!\!<\!\! \frac{N}{3}$ Byzantine replicas in each region.

\begin{lemma}
If a correct client broadcasts request req, then every correct replica eventually receives req and puts it in the respective shard.
\end{lemma}
\begin{proof}
Due to our system's synchronous model assumption that holds after GST, the correct client periodically sends the request to all replicas until the request is committed.
\end{proof}

\begin{lemma}
After GST, all correct replicas across clusters start executing the common case in round r before t, and no correct replica enters a round change before t + $\triangle$, every correct replica commits req. 
\end{lemma}
\begin{proof}
Let $\delta$ be the upper bound on the message delay after GST and let a correct leader propose a request req before t. All correct replicas across clusters receive at least N-F commit messages for r before t + 11$\delta$(See Figure~\ref{GeoNewBigBFT}). All correct replicas will accept these messages since they all enter round r by time t. As every correct replica receives at least N-F commits, every correct replica commits r by t + 11$\delta$. Therefore, $\triangle$=11$\delta$.
\end{proof} 

\begin{theorem}
Following from lemmas 3 and 5, the correct client request req will be broadcasted and committed by all honest replicas in all clusters.
\end{theorem} We defer the proof of the Theorem to the Appendix.

%% file: Evaluation.tex
\section{Implementation and Evaluation}
\label{sec:Evaluation}
\subsection{Implementation and setup}
\begin{figure}[!h]
	\centering
	\vspace*{-4mm}
	\includegraphics[width=3.5in,height=1.8in]{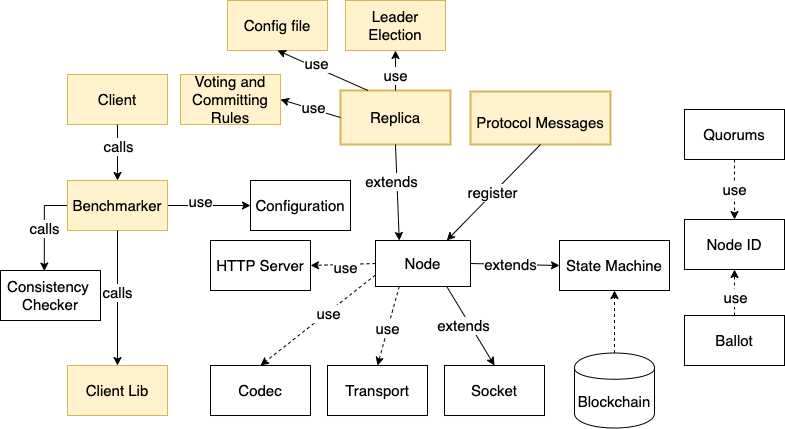}
	\caption{\textbf{The PaxiBFT architecture}}
	\vspace*{-4mm}
	\label{fig:Paxibft}
\end{figure}

We build BunchBFT, MirBFT, and BunchBFT-X from the open-source prototype of PaxiBFT~\cite{PaxiBFT}, written in GoLang. PaxiBFT is for prototyping, evaluating, and benchmarking BFT consensus and replication protocols. PaxiBFT uses core network files from Paxi~\cite{PAXI}. As shown in Figure~\ref{fig:Paxibft}, PaxiBFT readily provides most functionality that any coordination protocol needs for replication protocols. 

Our experiments are performed on up to 25 m5a.large virtual machines(VM), each of which has 2 vCPU, 8GiB RAM, and 10Gbps network throughput. Those VMs connect to Wide Area Network(WAN) across 5 AWS regions(Ohio, N.California, Oregon, N.Virginia, and Canada). To ensure that the client performance does not impact the results, we used the larger m5a.xlarge instances with 4 vCPUs for the clients. Finally, we used a message with and without a payload size equal to 128 bytes and 128K respectively.

\subsection{Evaluation}

\begin{figure}[!h]
	\centering
	\vspace*{-2mm}
	\includegraphics[width=3.5in,height=1.8in]{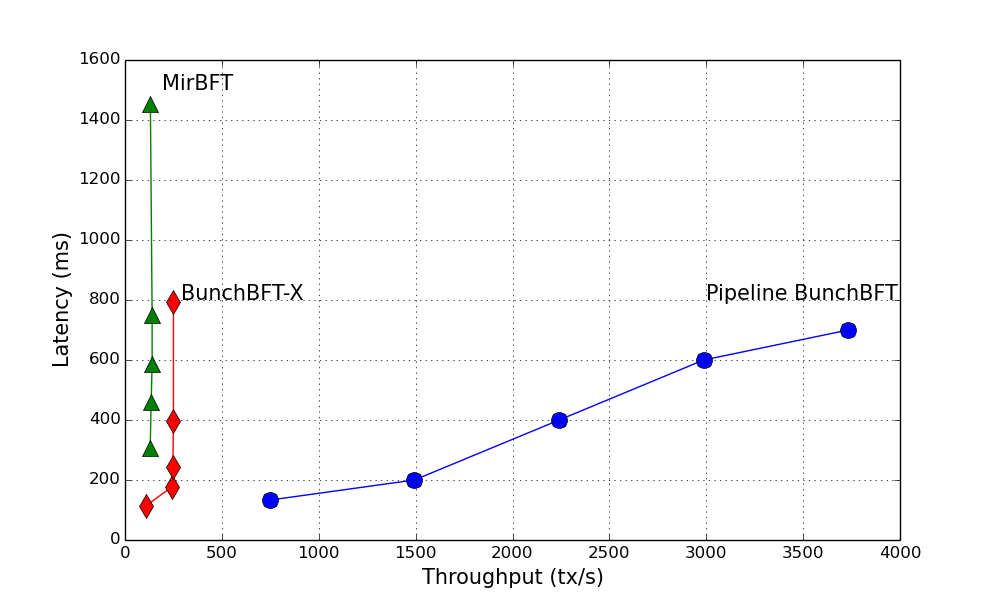}
	\caption{\textbf{Latency and Throughput on 25-node WAN cluster}}
	\vspace*{-2mm}
	\label{lvst}
\end{figure}

We evaluate the performance of our BunchBFT and BunchBFT-X prototype protocols and compare them with MirBFT~\cite{Mir}. To push system throughput, we varied the number of clients and used different message sizes. Figure~\ref{lvst} shows the scalability using 128 bytes messages. As the figure shows, BunchBFT achieves the best scalability since it allows all clusters to work in parallel serving more transactions. It scales up to 200 clients reaching a peak of 3750 Tx/s. On the other hand, BunchBFT-X scales to 200 clients with a peak throughput equal to 250 Tx/s. This is reasonable because BunchBFT-X is limited by its cross-cluster transactions. In BunchBFT-X experiment, the transactions across all available clusters capture the worst-case scenario of BunchBFT-X's throughput. BunchBFT-X’s throughput will increase significantly with one or more parallel cluster transactions. MirBFT scales to 200 clients, however, its throughput is 150 Tx/s. This refers to the extensive load of messaging on the replicas that the PBFT messaging pattern imposes. 

We defined latency is the time taken for the transaction to be approved and published. In Figure~\ref{lvst}, BunchBFT's latency is small with a single cluster, but it increases as the network clusters grow. On the other hand, the latency of BunchBFT-X and MirBFT increases rapidly with the increase in network size. MirBFT latency is higher than both BunchBFT-X and BunchBFT due to geographical properties depending on the region of request origin. 

\begin{figure}[!h]
	\centering
	\vspace*{-2mm}
	\includegraphics[width=3.5in,height=1.8in]{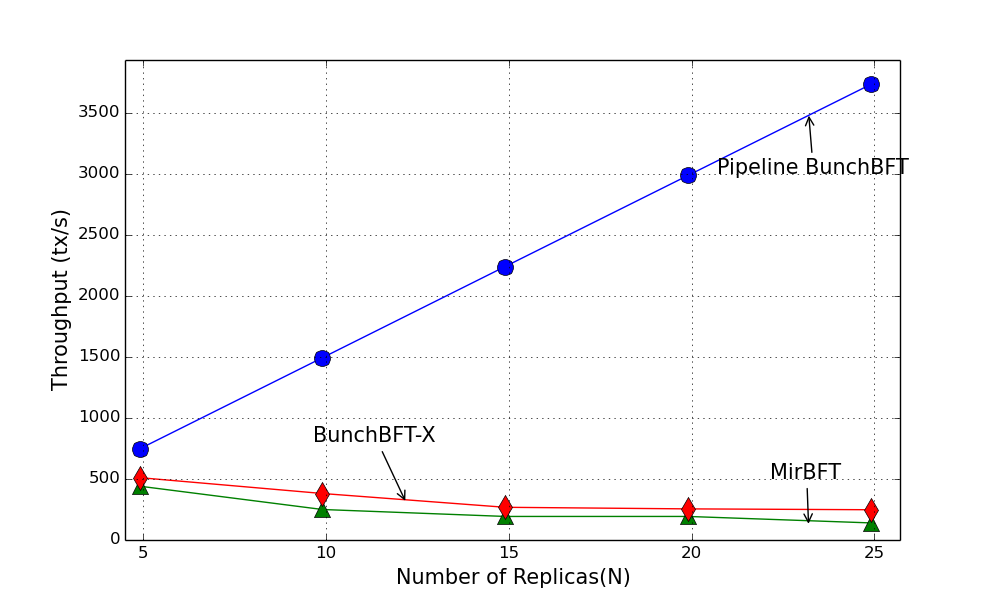}
	\caption{\textbf{System's Throughput on 25-node WAN cluster}}
	\vspace*{-2mm}
	\label{Thr1}
\end{figure}

\begin{figure}[!h]
	\centering
	\vspace*{-2mm}
	\includegraphics[width=3.5in,height=1.8in]{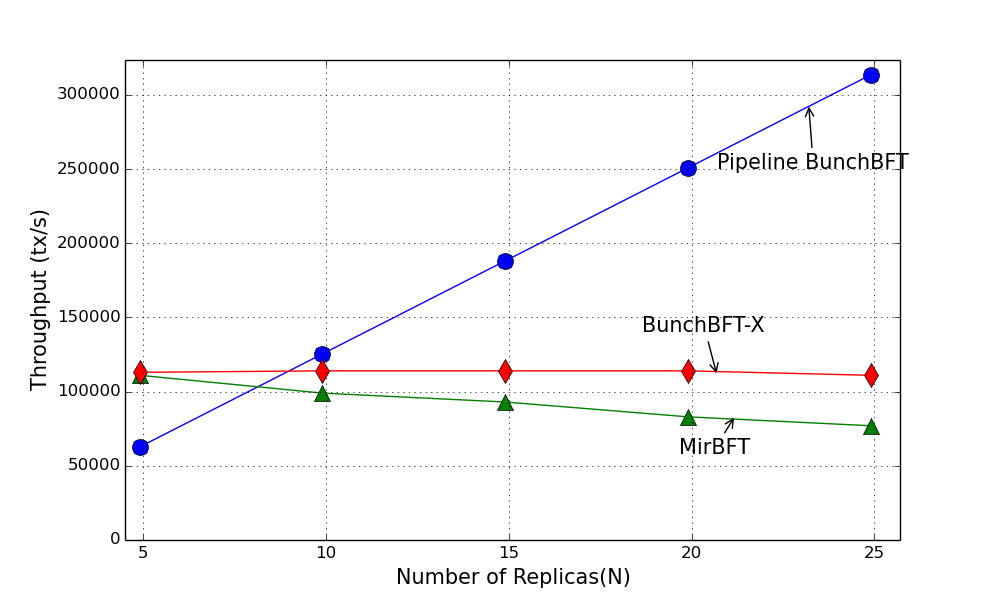}
	\caption{\textbf{System's Throughput with a block size equals to 128 KB on 25-node WAN cluster}}
	\vspace*{-2mm}
	\label{Thr2}
\end{figure}

We also measure the throughput, which is the number of committed transactions per second that are appended to the blockchain(Figure~\ref{Thr1},~\ref{Thr2}). The figures show the system's throughput using messages equal to 128 bytes and 128K respectively. The BunchBFT scales with the increasing number of replicas. On the other hand, BunchBFT-X and MirBFT degrade with the increase in network size.

%% file: Conclusions.tex
\section{Conclusions}
\label{sec:conc}
In this paper, we introduced a new BFT protocol called BunchBFT where we improved the scalability significantly by using a new cluster-based design. BunchBFT achieves 10X improved performance as compared to state-of-the-art MirBFT\cite{Mir}. Our experimental results show that BunchBFT scales to hundreds of clients in WAN, while the throughput of MirBFT drops quickly as the number of clients increases.